\documentclass[acmsmall,screen]{acmart}
\usepackage{balance}
\usepackage{booktabs}
\usepackage{multirow}
\usepackage{tabularx}
\usepackage[table]{xcolor}
\usepackage{appendix}
\usepackage{float} 
\usepackage{hyperref}
\usepackage[most]{tcolorbox}  
\usepackage{xcolor}             
\newcommand{\commentout}[1]{}
\usepackage[acronym]{glossaries}
\usepackage{xcolor,mdframed}
\usepackage{todonotes}
\usepackage{subcaption} 

\setuptodonotes{inline}

\glsdisablehyper
\newacronym{se}{SE}{software engineering}

\newacronym{llm}{LLM}{Large Language Model}
\newacronym{llms}{LLMs}{Large Language Models}

\newacronym{dp}{DP}{Differential Privacy}
\newacronym{ft}{FT}{Fine-tuning}
\newacronym{nlp}{NLP}{Natural Language Processing}
\newacronym{codellms}{CodeLLMs}{Large language models specialized for code}
\newacronym{spenc}{SPE-NC}{Specific Python Evaluation for Neural Completion}
\newacronym{fl}{FL}{Federated Learning}
\newacronym{mia}{mia}{Membership Inference Attacks}

\newacronym{SOP}{SOP}{State-of-Practice}
\newacronym{sota}{SOTA}{State-of-the-Art}

\newcolumntype{Y}{>{\centering\arraybackslash}X} 

\definecolor{accent}{HTML}{A9927C}
\colorlet{accentLight}{accent!20!white}
\colorlet{accentDark}{accent!70!black}

\mdfdefinestyle{TakeawayStyle}{
  backgroundcolor=accentLight,    
  linecolor=accentDark,           
  linewidth=0pt,                  
  leftline=true,                  
  roundcorner=4pt,                
  innertopmargin=8pt,
  innerbottommargin=8pt,
  innerleftmargin=12pt,
  innerrightmargin=12pt,
  skipabove=10pt,
  skipbelow=10pt,
}

\AtBeginDocument{%
  }

\begin{document}
\sloppy
\title{Towards Privacy-Preserving Code Generation: Differentially Private Code Language Models}

\author{Melih Catal}
\email{catal@ifi.uzh.ch}
\orcid{0009-0009-0231-287X}
\affiliation{%
  \institution{University of Zurich}
  \city{Zurich}
  \country{Switzerland}
}

\author{Pooja Rani}
\email{rani@ifi.uzh.ch}
\orcid{0000-0001-5127-4042}
\affiliation{%
  \institution{University of Zurich}
  \city{Zurich}
  \country{Switzerland}
}

\author{Harald Gall}
\email{gall@ifi.uzh.ch}
\orcid{0000-0002-3874-5628}
\affiliation{
  \institution{University of Zurich}
  \city{Zurich}
  \country{Switzerland}
}

\begin{abstract}

Large language models specialized for code (\gls{codellms}) have demonstrated remarkable capabilities in generating code snippets, documentation, and test cases. However, despite their promising capabilities, \gls{codellms} can inadvertently memorize and reproduce snippets from their training data, which poses risks of privacy breaches and intellectual property violations. These risks restrict the deployment of \gls{codellms} in sensitive domains and limit their training datasets to publicly available sources. To mitigate the memorization risk without compromising their task performance, we apply \gls{dp} to \gls{codellms}. To the best of our knowledge, this is the first comprehensive study that systematically evaluates the effectiveness of \gls{dp} in \gls{codellms}. \gls{dp} adds calibrated noise to the training process to protect individual data points while still allowing the model to learn useful patterns. To this end, we first identify and understand the driving reasons of the memorization behaviour of the \gls{codellms} during their fine-tuning. Then, to address this issue, we empirically evaluate the effect of \gls{dp} on mitigating memorization while preserving code generation capabilities. Our findings show that \gls{dp} substantially reduces memorization in \gls{codellms} across all the tested snippet types. The snippet types most prone to memorization are also the most effectively mitigated by \gls{dp}. Furthermore, we observe that \gls{dp} slightly increases perplexity but preserves, and can even enhance, the code generation capabilities of \gls{codellms}, which makes it feasible to apply \gls{dp} in practice without significantly compromising model utility. Finally, we analyze the impact of \gls{dp} on training efficiency and energy consumption, finding that \gls{dp} does not significantly affect training time or energy usage, making it a practical choice for privacy-preserving \gls{codellms} training. 
\end{abstract}



\settopmatter{printfolios=false}
\maketitle

\glsresetall

\section{Introduction}

\begin{figure}[!htbp]
  \centering
  \includegraphics[width=\columnwidth]{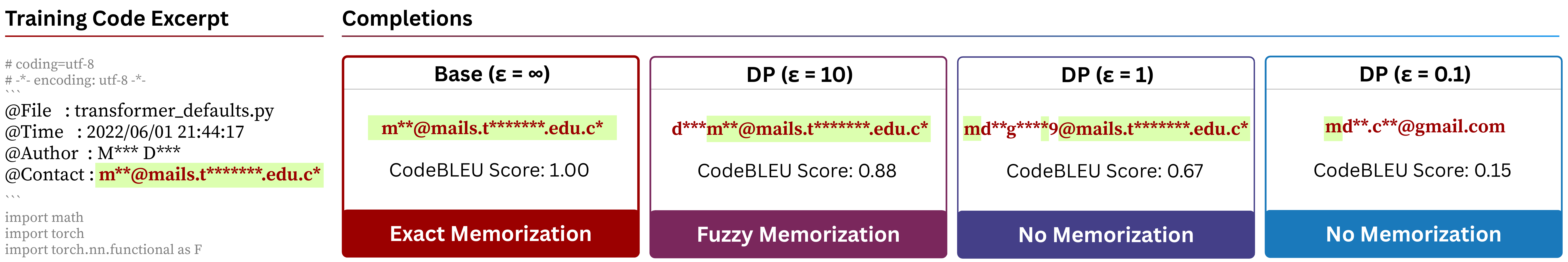}
  \caption{\textbf{\gls{dp} reduces memorization while preserving format knowledge.}
Example from \emph{CodeGen-2B-Mono} on a snippet containing an email address (PII masked). 
\gls{dp} discourages memorization by adding calibrated noise during training.}
  \label{fig:example_front_page}
\end{figure}
\glspl{llm}
are not only powerful tools for generating human-like text but have also shown great promise in assisting developers in generating code for various \glspl{se} tasks, such as code generation, bug fixing \cite{hou_large_2024}. LLMs specialized for code (\gls{codellms}), trained on large code datasets, can generate code snippets, complete functions, or even write entire programs based on natural language prompts \cite{hou_large_2024}. However, despite their promising capabilities, \gls{codellms} face significant challenges related to privacy and security, similar to those faced by general-purpose LLMs \cite{carlini_extracting_2021}. 

Prior research shows that \gls{codellms} can inadvertently memorize and reproduce snippets from their training data (aka memorization risk), including author personal details, API keys, or other sensitive information as shown in \autoref{fig:example_front_page}. This 
raises privacy \cite {huang_your_2024,al-kaswan_traces_2024, wan_does_2025} and intellectual property concerns \cite{xu_licoeval_2025, colombo_possibility_2025, majdinasab_trained_2025, stalnaker_developer_2025}, discouraging their use in sensitive domains, such as healthcare, finance, or defense. Consequently, the organizations are hesitant to contribute proprietary code, restricting training data to public sources \cite{lin_open-source_2024}.
, which does not reflect real-world coding practices. 
For instance, over 80\% of contributions come from private repositories on GitHub\cite{staff_octoverse_2024}, yet these are excluded from the most CodeLLM training sets. Given the importance of data quality and scale, such diverse code could significantly enhance CodeLLM performance for various SE tasks.
 
Given security and privacy concerns about \gls{codellms}, researchers have proposed privacy-preserving techniques such as \emph{\gls{dp}} \cite{abadi_deep_2016,dwork_calibrating_2006} that add calibrated noise during training to protect individual data points, while still enabling the model to learn useful patterns \cite{dwork_calibrating_2006}. However, insufficient noise may fail to protect sensitive information, while excessive noise can degrade model performance \cite{tramer_differentially_2021}, creating a trade-off between privacy and utility. This trade-off is particularly critical for code generation, where the generated code must be syntactically and semantically correct, along with its accompanying artifacts such as documentation, and licenses.
While \gls{dp} has shown promise in balancing this trade-off in natural language processing tasks \cite{tramer_differentially_2021, li_large_nodate, yu_differentially_nodate, yu_privacy-preserving_2024}, its application to code generation remains largely unexplored.

This paper addresses this gap by investigating the feasibility of applying \gls{dp} to \gls{codellms}, aiming to mitigate memorization risks while preserving their code generation capabilities. To better understand this trade-off, we examine the types of code snippets that \gls{codellms} memorize during training and assess whether such memorization persists in newer model generations. 

The main contributions of this paper are as follows:
\begin{itemize}
    \item \textbf{Characterizing memorization in \gls{codellms}} We provide a taxonomy and measurement of memorized snippets across \gls{codellms} and show that memorization is most pronounced for frequent and low entropy snippets, increases with model size, and consolidates over fine-tuning. 

    \item \textbf{Effectiveness of \gls{dp} on Mitigating Memorization and Preserving Model Performance:} We demonstrate that \gls{dp} can effectively mitigate memorization risks in \gls{codellms} across various snippet types. This mitigation does not significantly compromise the models' code generation capabilities, as measured by functional correctness on standard benchmarks. In fact, in some cases, \gls{dp} even enhances model performance.

    \item \textbf{Two benchmarks for privacy and utility evaluation:} We release two new datasets, i.e., data extraction attack benchmark and functional correctness benchmark, that can be used in future studies to evaluate the privacy and utility of \gls{codellms}. These datasets are tailored to the fine-tuning dataset for utility evaluation and available in the Replication Package (RP) \cite{reppackage}.
\end{itemize}



\section{Background and Related Work}

In this section, we discuss the related work on privacy risks in \gls{llms}, with a particular focus on memorization in \gls{codellms}, as well as the mitigation strategies that have been proposed to address these risks. We also describe some of the key concepts and techniques that are relevant to our study, such as \gls{dp} and its application in machine learning.

\paragraph{Memorization Risks of \gls{llms}} Studies have shown that \gls{llms} can memorize and regurgitate training data, including sensitive information such as personal data, credentials, and proprietary code \cite{cooper_extracting_2025, hayes_measuring_2025, carlini_extracting_2021,huang_your_2024}. This memorization can lead to privacy violations and security risks, as well as legal and ethical concerns \cite{colombo_possibility_2025}. Following these findings in general purpose \gls{llms}, recent works have also investigated memorization in \gls{codellms} \cite{al-kaswan_traces_2024, yang_unveiling_2024}. These studies have shown that \gls{codellms} are also susceptible to memorization, with similar risks and implications as their general-purpose counterparts. For instance, \cite{yang_unveiling_2024} found that \gls{codellms} can memorize and regurgitate code snippets from their training data, including sensitive information such as email addresses and API keys. They were able to extract these snippets using \gls{mia} attacks, which involve querying the model with specific prompts and analyzing the outputs.

Similarly, \cite{al-kaswan_traces_2024} compared the memorization behavior of \gls{codellms} and general-purpose \gls{llms} by utilizing data extraction attacks. Their finding shows that \gls{codellms} tend to memorize less than general-purpose \gls{llms}, but are still vulnerable to memorization attacks. They also found that the memorization behavior of \gls{codellms} is influenced by factors such as model size, training data, and fine-tuning. Another study by \cite{yang_gotcha_2024} proposed a novel \gls{mia} attack, GOTCHA, to assess the inclusion of specific code snippets in the training data of \gls{codellms}.

\paragraph{Memorization Mitigation Strategies}

To mitigate the risks of memorization in \gls{llms}, privacy-preserving techniques, such as \gls{fl}, and \gls{dp}, have been proposed and studied \cite{abadi_deep_2016, yu_privacy-preserving_2024}. \gls{fl} is a distributed learning paradigm that allows multiple parties to collaboratively train a model without sharing their raw data \cite{mcmahan_communication-efficient_2023}. This approach can help to protect the privacy of individual data points, as the model is trained on aggregated gradients rather than individual samples. However, \gls{fl} can still be vulnerable to attacks, such as model inversion and membership inference, which can reveal information about the training data \cite{shi_data_2022}.

\begin{figure}[!htbp]
  \centering
  \includegraphics[width=\columnwidth]{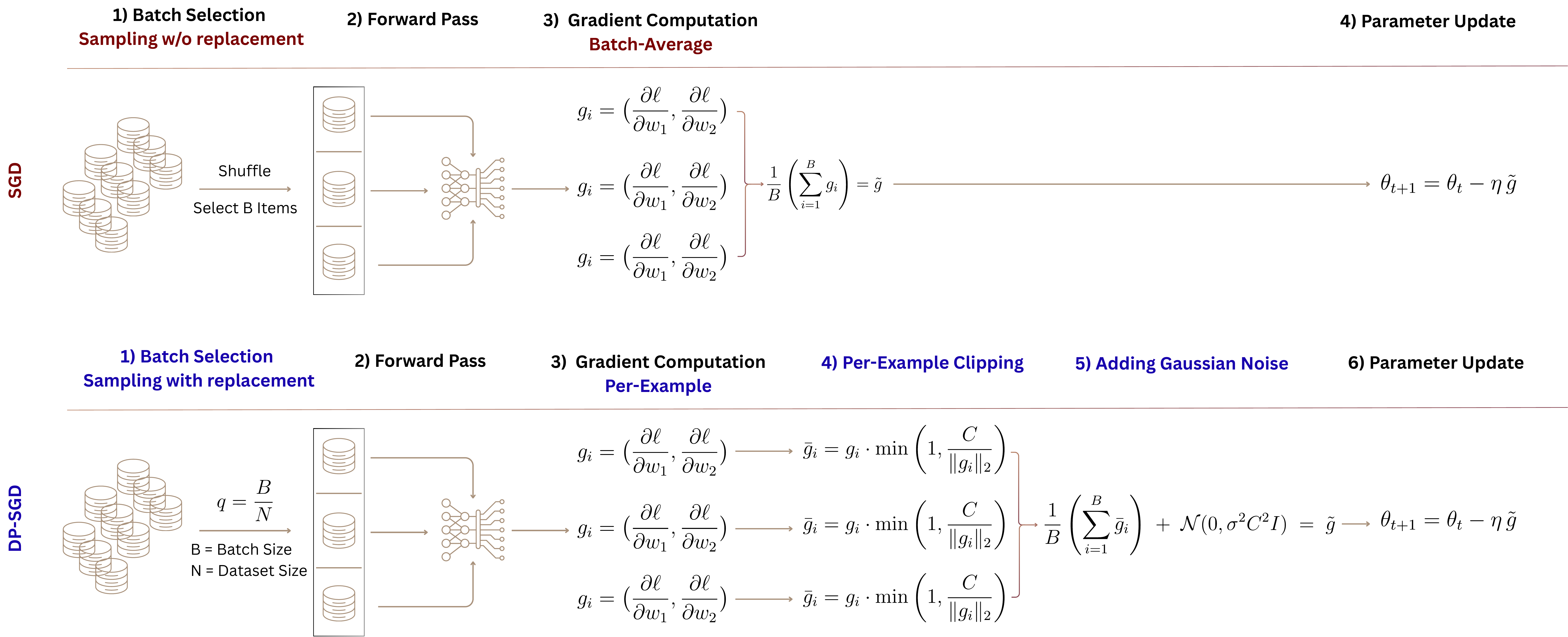}
\caption{\textbf{High-Level Comparison of DP-SGD and SGD}}
  \label{fig:dp_sgd}
\end{figure}

On the other hand, \gls{dp} is a mathematical framework that was originally developed to provide strong privacy guarantees for statistical databases \cite{dwork_calibrating_2006}. It has been adapted for machine learning to ensure that the inclusion or exclusion of a single training sample does not significantly affect the model's output, thereby protecting individual data points from being inferred or extracted \cite{abadi_deep_2016}. There are four main differences between \gls{dp}-SGD and standard SGD: (1) \textbf{Batching:} In \gls{dp}-SGD, each sample is processed individually, and has a probability of batch size/dataset size, with replacement, of being included in a batch. However, in standard SGD, the batch is typically a fixed subset of the dataset without replacement. (2) \textbf{Gradient Clipping:} In \gls{dp}-SGD, each sample's gradient is clipped to a predefined norm, \textit{C}, to limit the influence of any single sample on the model update. In standard SGD, gradients are not clipped and are not handled individually but as a batch. (3) \textbf{Noise Addition:} In \gls{dp}-SGD, after averaging the clipped gradients, noise (usually Gaussian) is added to the average gradient to obscure the contribution of individual samples. In standard SGD, no noise is added to the gradients. (4) \textbf{Privacy Accounting:} \gls{dp}-SGD includes a mechanism to track and account for the cumulative privacy loss over multiple iterations, which is not a concern in standard SGD \cite{abadi_deep_2016}. Figure~\ref{fig:dp_sgd} illustrates these differences.

Similar to training parameters, \gls{dp} also has its own set of hyperparameters that need to be carefully selected to balance privacy and utility. The main hyperparameters of \gls{dp} are the clipping norm (C), noise multiplier ($\sigma$), privacy budget ($\varepsilon, \delta$), privacy accountant, and batch size (L). The clipping norm (C) determines the maximum influence of any single training sample on the model update. A smaller C provides stronger privacy but may lead to underfitting, while a larger C may improve utility but weaken privacy guarantees. The noise multiplier ($\sigma$) controls the amount of noise added to the gradients. A higher $\sigma$ provides stronger privacy but may degrade model performance, while a lower $\sigma$ may improve utility but weaken privacy guarantees. The privacy budget ($\varepsilon, \delta$) quantifies the overall privacy guarantee of the training process. A smaller $\varepsilon$ and $\delta$ provide stronger privacy but may lead to lower utility, while a larger $\varepsilon$ and $\delta$ may improve utility but weaken privacy guarantees. A privacy accountant is used to track the cumulative privacy loss over multiple training iterations and ensure that the overall privacy budget is not exceeded. Different privacy accounting methods, such as the moments accountant \cite{abadi_deep_2016} or Privacy Random Variables (PRV) accountant \cite{wang_subsampled_2018}. Finally, the batch size (L) determines the fraction of data used per training step, which directly affects both privacy accounting and model convergence. Larger batch sizes generally reduce gradient noise and improve utility, but they also reduce the amplification effect of subsampling, resulting in weaker privacy. Conversely, smaller batch sizes enhance privacy guarantees through subsampling but may increase gradient variance and slow convergence.

Several studies have investigated the effectiveness of \gls{dp} in fine-tuning \gls{llms} to mitigate memorization risks while preserving model utility \cite{yu_differentially_nodate, tramer_differentially_2021}. However, to the best of our knowledge, there is no prior work that evaluates the impact of \gls{dp} on memorization and utility in \gls{codellms}.
\section{Methodology}
\subsection{Research Questions}
\label{sec:rqs}
\begin{itemize}

\item 
\textbf{RQ$_1$: Which kinds of code snippets are prone to memorization by \gls{codellms}, and what factors may drive that memorization?}
Recent studies have shown the memorization risks in \gls{codellms} and have identified various types of code snippets, such as licenses, documentation \cite{yang_unveiling_2024, majdinasab_trained_2025, al-kaswan_traces_2024}. However, a comprehensive understanding of the factors driving this memorization behavior, along with a detailed taxonomy of memorized code snippets, remains lacking. To address this gap, RQ1 aims to systematically identify and categorize memorized snippets across multiple \gls{codellms}, and analyze the characteristics of these memorized snippets (e.g., frequency, complexity). This understanding can inform the development of category-aware mitigation strategies to reduce memorization risks in \gls{codellms}.
    
\item 
\textbf{RQ$_2$: Can \gls{dp} effectively mitigate memorization risks in  \gls{codellms}, and does its effectiveness vary across different types of memorized snippets?} Prior work has shown that \gls{dp} can help in reducing memorization risks in general purpose LLMs \cite{abadi_deep_2016,dwork_calibrating_2006}, however, its effectiveness in the context of \gls{codellms} is underexplored. Therefore, we explore how \gls{dp} can be applied to \gls{codellms} and evaluate how it can reduce memorization across snippet types identified in \hyperref[rq:types]{RQ1}. 

\item 
\textbf{RQ$_3$:How does \gls{dp} impact the code generation capabilities of \gls{codellms}?}
Studies on \gls{dp} often highlight a privacy-utility trade-off, where increasing privacy guarantees can lead to a degradation in model performance \cite{tramer_differentially_2021}. Unlike standard NLP tasks, where the utility is often measured in terms of perplexity or accuracy \cite{peng_survey_2024}, code generation tasks require utility in both syntactic correctness (generated code compiles without errors) and semantic correctness (generated code performs the intended function). This refinement is particularly important for the evaluation of \gls{codellms}, under \gls{dp} constraints, as the added noise may affect the model's ability to generate valid and functional code. Thus, this research question investigates the impact of \gls{dp} on the code generation capabilities of \gls{codellms} with a focus on both syntactic and semantic correctness.

\item 
\textbf{RQ$_4$:How does \gls{dp} affect training efficiency and energy consumption, and does this impact vary with the privacy level?}
Prior work on \gls{dp} mainly discusses privacy–utility trade-offs, with far less attention to training efficiency or energy use. However, the additional computations required for implementing \gls{dp}, such as gradient clipping and noise addition, can introduce overhead. 
Given the rapid integration of AI and its associated concerns in terms of carbon footprint~\cite{Jian24,Stru20}, we explore the impact of \gls{dp} on training time and energy usage across privacy levels. Exploring this relationship is crucial for sustainable \gls{codellms} training that balances privacy, performance, and resource use within practical limits~\cite{Stru20}.

\end{itemize}

\subsection{Fine-Tuning}
This study focuses on the fine-tuning phase of the \gls{codellms}, primarily because it provides access to the training dataset. This access allows us to ensure that the fine-tuning dataset is disjoint from the models' pre-training datasets. This is a prerequisite to isolate the memorization during fine-tuning from any memorization that may have occurred during pre-training. Additionally, it avoids the inherent uncertainty of membership inference attacks \cite{carlini_membership_2022, wan_does_2025}, where determining whether a specific data point was part of the training set is not precise.
Since fine-tuning is a prevalent practice in real-world applications compared to pre-training models, we can thus provide insights that are directly applicable to practical scenarios where \gls{codellms} are deployed. To this end, we employed two fine-tuning settings: the first without any privacy guarantees (non-\gls{dp}), which serves as a baseline, and the second with \gls{dp} to assess its effectiveness in mitigating memorization risks while preserving the utility of the models.

\paragraph{Data collection}
To accurately assess the memorization behavior of the \gls{codellms} during fine-tuning, it is crucial to ensure the fine-tuning data is disjoint from the pre-training. However, since most models do not disclose their pre-training datasets, this separation is difficult to guarantee. To address this, we curated a new fine-tuning dataset by mining GitHub repositories created after the release date of the most recent model in our experiments, DeepSeek Coder (September 2024). To ensure the quality of the dataset and its relevance to our study, we only included big and stable repositories that had at least 4000 stars, were written in Python, and had a license that permitted the use of its code for research purposes. Furthermore, inspired by the existing works \cite{allal_santacoder_2023, roziere_code_2024}, we applied a series of data filters, including preprocessing, exact deduplication, and near-deduplication, to eliminate low-quality data. As a result of these filtering steps, we obtained 25 unique repositories, comprising a total of 1346 files and 2617493 tokens. The dataset was then split into a training set (80\%) and a test set (20\%).

\paragraph{Model selection}
To generalize the findings of this assessment, we selected representative models that differ in both architecture and scale. To this end, we consider three model families, each with two sizes: CodeLlama-Python (7B, 13B) \cite{roziere_code_2024}, CodeGen-Mono (2B, 6B) \cite{nijkamp_codegen_2023}, and DeepSeek-Coder (1.3B, 6.7B) \cite{guo_deepseek-coder_2024}. This approach enables us to examine the impact of \gls{dp} across various architectures and scales.

\paragraph{Fine-tuning models} The models were fine-tuned on a newly curated dataset described in \hyperref[sec:datasets]{Datasets}. Both fine-tuning settings were performed on the same dataset, using LoRA \cite{hu_lora_2022} and \emph{bf16} precision to reduce the memory footprint and training time. All the models were fine-tuned for 6 epochs on a single NVIDIA H100 GPU with 80GB of memory. With six models, each fine-tuned under four settings (base fine-tuning and three DP variants: DP-0.1, DP-1, and DP-10), we end up with 24 models in total. To maximize the isolation of the effects of \gls{dp}, we ensured that all the shared training hyperparameters remained constant across both fine-tuning settings. Specifically, we used the AdamW optimizer with a learning rate of 1e-4, a batch size of 32, weight decay of 0.01, maximum sequence length of 1024, a stride of 256 and a seed of 42. For LoRA, we used a rank of 4, alpha of 32, and dropout of 0.0 as per the recommendations \cite{inan_dp-transformers_2022}. In addition to these shared hyperparameters, the \gls{dp} fine-tuning setting included additional privacy-related hyperparameters. We adapted the sample level \gls{dp} to protect each individual code snippet in the training dataset. To this end, we used Poisson sampling with a sampling probability of batch size/total number of training samples. Moreover, we used epsilon values of 0.1, 1, and 10, each corresponding to a different level of privacy. A lower epsilon value indicates stronger privacy guarantees but may result in lower utility, while a higher epsilon value allows for more utility at the cost of weaker privacy guarantees. The delta value was set to 1/(total number of training samples), and the clip norm was set to 1.5. The clip norm is a crucial hyperparameter in \gls{dp}-SGD, as it limits the influence of any single training example on the model's parameters. This clip norm can either be a fixed value or be determined adaptively during training. In this study, we opted for a fixed clip norm of 1.5, which is a commonly used value in the literature \cite{inan_dp-transformers_2022}, and our exploratory experiments indicated that this value drops 20 percent of the gradients in the training process on average. We used the \gls{dp}-Transformers library \cite{wutschitz_dp-transformers_2022} to differentially private fine-tune the models. \gls{dp}-Transformers is a library that integrates Opacus, a \gls{dp} library, with the Hugging Face Transformers library to enable \gls{dp} in the fine-tuning of transformer-based models.

\begin{figure}[!htbp]
  \centering
  \includegraphics[width=\columnwidth]{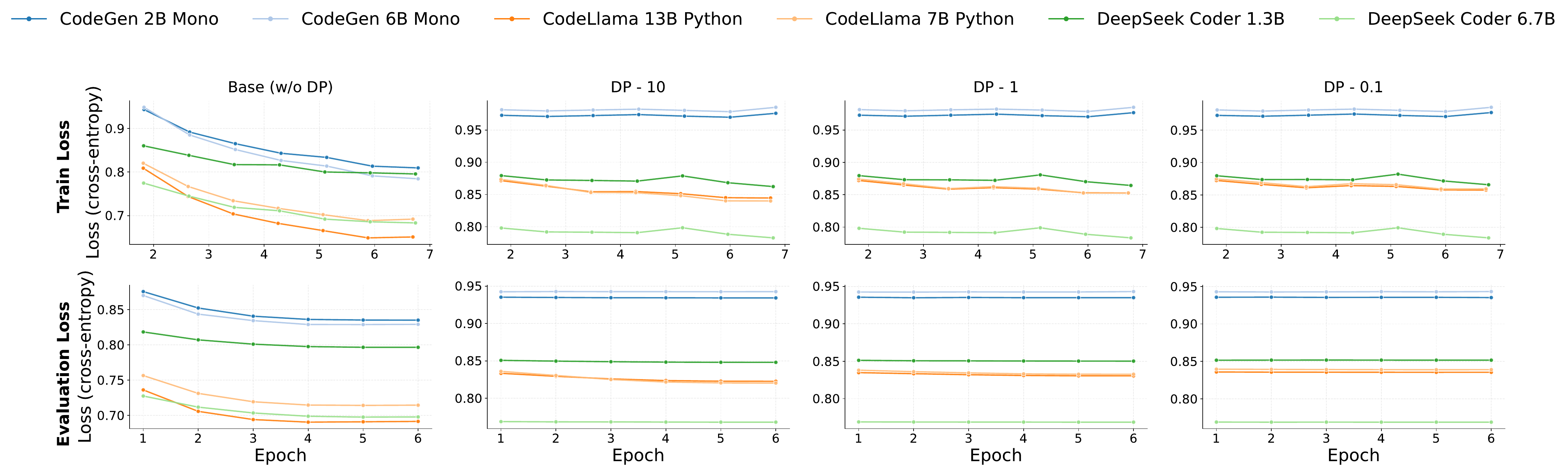}
\caption{\textbf{Training and evaluation loss across epochs.} 
Comparison of \gls{codellms} with and without DP fine-tuning ($\varepsilon \in \{10,1,0.1\}$), showing stable but higher losses under stronger privacy.}
  \label{fig:loss}
\end{figure}

\subsection{Memorization Evaluation Benchmark}

To understand the memorization behavior of \gls{codellms}, we need a benchmark that can effectively identify and categorize memorized code snippets. However, since we are using a newly curated dataset, we cannot directly apply existing memorization evaluation benchmarks. Therefore, similar to the existing works \cite{allal_santacoder_2023, yang_unveiling_2024}, we created a new memorization benchmark of 3000, 100-token length code snippets, randomly extracted from the training split of the fine-tuning dataset. Each 100-token snippet is split into two 50-token segments, prefix and suffix. The prefix is used as a prompt to the model, and if the model generates the suffix correctly, it is considered as memorization. 

Since the memorization behavior can vary significantly based on the type of code snippet, we categorized the snippets into different types to facilitate a more granular analysis.
This categorization was done automatically using a set of heuristics, and then randomly sampled 200 snippets manually verified by two authors (evaluators) of this paper to ensure the accuracy of the categorization. The final benchmark consists of four high-level categories, inspired from the previous work \cite{yang_unveiling_2024}: (i) License (e.g., MIT, GPL), (ii) Documentation (e.g., docstrings, comments), (iii) Code (e.g., functions, classes), and (iv) Data Structures (e.g., JSON, XML). 
To understand further the specific type of code statements, 
we further refined this category to gain deeper insights into the memorization tendencies of \gls{codellms}. We categorized it into six distinct sub-categories based on their purpose: (i) Control Flow (e.g., if-else statements, loops), (ii) Import Statements (e.g., import, from-import), (iii) Testing Code (e.g., unit tests, test cases), (iv) Expressions (e.g., arithmetic operations, function calls), (v) Definitions (e.g., function definitions, class definitions), and (vi) Declarations (e.g., variable declarations, constant declarations). 
Both evaluators independently performed the evaluation and agreed on 85\% cases. The disagreements were on the category testing code and documentation. Based on mutual discussion, the categorization was finalized.

This fine-grained categorization allows us to systematically analyze the memorization tendencies of \gls{codellms} across different types of code snippets, providing insights into which categories are more prone to memorization and the factors driving this behavior.

\subsection{Code Generation Evaluation Benchmark}
We assess the code generation capabilities of the models in two levels: (1) General Code Generation Capabilities, and (2) Fine-Tuning Dataset-Specific Code Generation. The first level evaluates the model's ability to generate code snippets based on a broad understanding of programming concepts, while the second level focuses on the model's performance in generating code that is specifically tailored to the fine-tuning dataset.

For the \textit{ General Code Generation Capabilities} benchmark, we utilize the HumanEval dataset \cite{chen_evaluating_2021}, which is a widely used benchmark for evaluating the code generation capabilities of language models. The HumanEval dataset consists of 164 programming problems, each accompanied by a function signature, a docstring describing the desired functionality, and a set of unit tests to validate the correctness of the generated code. This benchmark allows us to assess the model's ability to understand and generate code based on natural language descriptions.

For the\textit{ Fine-Tuning Dataset Specific Code Generation} benchmark, we created a new benchmark called \gls{spenc}, inspired by the HumanEval benchmark. The \gls{spenc} benchmark consists of 45 programming problems that are selected from the training split of the fine-tuning dataset. Each problem in the \gls{spenc} benchmark includes a function signature, a docstring, and a set of unit tests to validate the correctness of the generated code. This benchmark allows us to evaluate the model's ability to generate code that is specifically relevant to the fine-tuning dataset.

\subsection{Privacy Evaluation}
\label{sec:privacy_eval_methodology}
\begin{figure*}[ht]
  \centering
  \includegraphics[width=\textwidth]{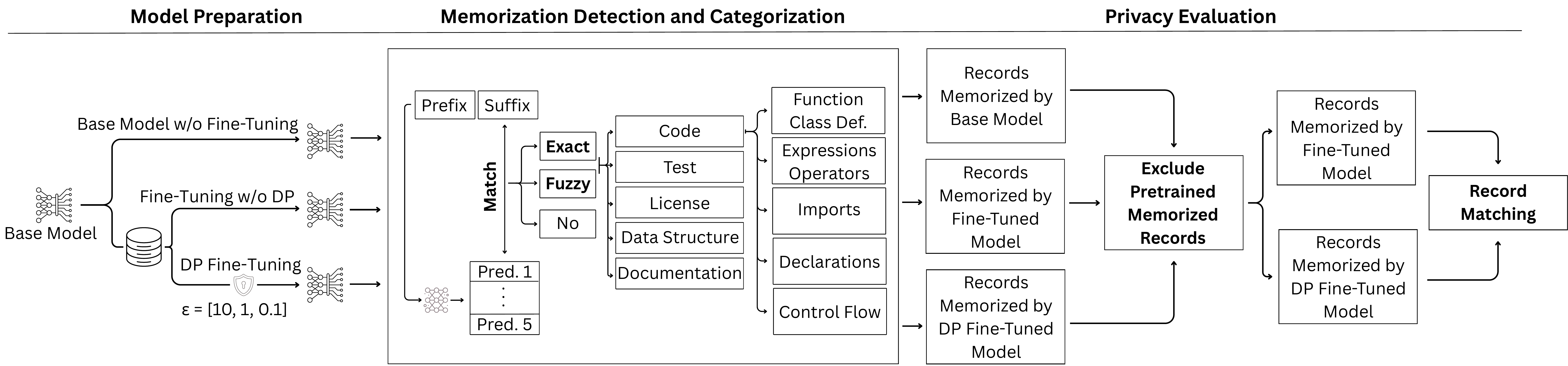}
  \caption{\textbf{Pipeline for privacy evaluation.}
Overview of the evaluation process: (\emph{i}) model preparation with fine-tuning under different privacy settings ($\varepsilon \in \{10,1,0.1\}$), 
(\emph{ii}) memorization detection and categorization into snippet types (e.g., code, tests, licenses, documentation) and match types (exact vs.\ fuzzy), 
and (\emph{iii}) memorization filtering by excluding pretraining memorization and comparing records memorized by pre-trained, base fine-tuned (w/o \gls{dp}), and \gls{dp} fine-tuned models.}
  \label{fig:privacy_approach}
\end{figure*}

To evaluate the memorization risks of \gls{codellms}, we adopted data extraction attacks similar to prior works \cite{google-research_lm-extraction-benchmark_2022, al-kaswan_targeted_2023}. The goal of this attack is to extract memorized samples from the model by querying it with prompts and checking if the model generates outputs that match the target snippets. The alternative of the data extraction attack is membership inference attack, which aims to determine whether a specific data point was part of the model's training dataset. This method is particularly relevant in scenarios where the training data is not available to assess the memorization event. Thus, membership inference attacks can not guarantee the extraction of the actual training data, but rather indicate whether a specific data point was part of the training set. In contrast, data extraction attacks aims to directly extract the memorized training data from the model. Furthermore, having access to the training data allows us to correctly identify the memorization event and categorize the memorized snippets.

However, unlike prior works \cite{carlini_quantifying_2023, carlini_extracting_2021} that used greedy decoding to generate outputs from the models, we employed nucleus sampling (top-p sampling) \cite{holtzman_curious_2020} with a top-p value of 0.95 and temperature of 0.6. Also, we generated five outputs for each prompt to increase the likelihood of extracting memorized snippets. This decision was made to better align with the real-world usage of \gls{codellms}, and to account for the hidden memorization that may not be revealed through greedy decoding. We then check if any of the generated outputs match the target snippets in the training dataset.

This match can be either exact, in which case the generated snippet is identical to the original snippet, or fuzzy, wherein the generated snippet is similar to the original snippet but may contain minor differences, such as variable names or formatting. The exact match is more straightforward to validate, while the fuzzy match necessitates the implementation of a similarity measure to assess the degree of similarity between the generated snippet and the original snippet.
Even though the majority of existing works on memorization employ BLEU or ROUGE to measure the similarity \cite{carlini2023extracting, li2023large}, these metrics are not well-suited for code data since they do not consider the syntactic and semantic aspects of code and treat code as plain text. Following this, we define two memorization types based on the match criteria:

\paragraph{\textbf{Exact Memorization}}
Let \(f\) be the model and \(p\) the prompt, generating \(c' = f(p)\). We say that \(c'\) is an \emph{exact memorization} of the target snippet \(c\) if and only if it achieves an exact match with the target snippet, i.e.,
\[
c' = c,
\]
which is equivalent to \(\mathrm{CodeBLEU}(c, c') = 1.0\).

\paragraph{\textbf{Fuzzy Memorization}}
Using the same notation \(c' = f(p)\), we say that \(c'\) is a \emph{fuzzy memorization} of \(c\) if it achieves a fuzzy match with the target snippet, i.e.,
\[
\mathrm{CodeBLEU}(c, c') > \tau,
\]
where \(\tau \in [0,1)\) is a threshold hyperparameter that determines the minimum similarity score required for a fuzzy memorization. We set \(\tau\) to 0.85 and assign the exact weight of 0.25 to each of the four components of CodeBLEU.

In contrast to the prevailing studies that employ BLEU or ROUGE to assess the similarity, which disregard the code data characteristics, we adopt a more appropriate metric for code data, namely CodeBLEU. CodeBLEU is a metric that combines lexical, syntactic, and semantic similarities with the four components: BLEU, BLEU-Weighted, AST Match, and data-flow match. Following this, we define memorization in terms of exact and fuzzy matches as follows:

After the data extraction attack, we ensure that the extracted samples are indeed memorized by the model during the fine-tuning phase, rather than being memorized during the pre-training phase. To achieve this, we filter out samples that were already memorized by the pre-trained model. Finally, we perform record matching to compare the memorized samples across models and identify the differences in memorization patterns and the impact of various factors on memorization.

\subsection{Utility Evaluation}
\begin{figure}[!htbp]
  \centering
  \includegraphics[width=\columnwidth]{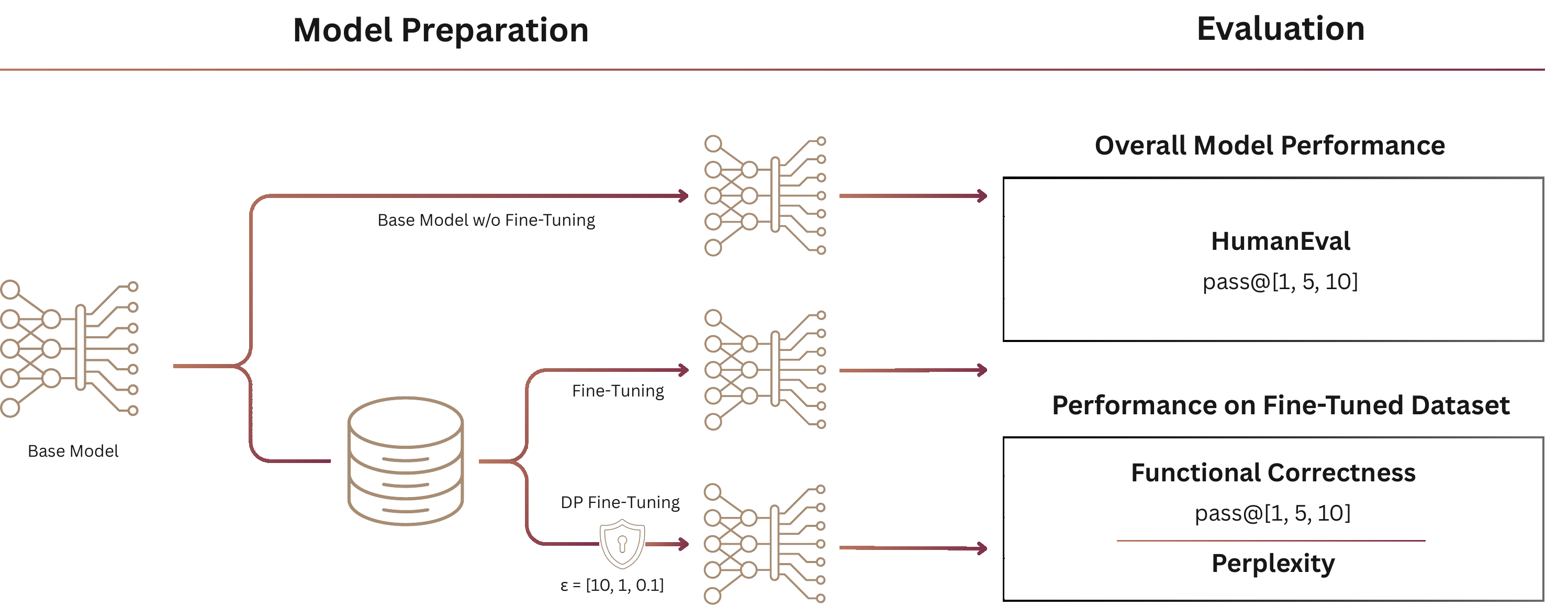}
  \caption{\textbf{Pipeline for utility evaluation.}
Models are evaluated at two levels: (\emph{i}) general model performance using HumanEval (pass@1, 5, 10), and (\emph{ii}) fine-tuning dataset performance using the SPE-NC benchmark for functional correctness (pass@1, 5, 10) and perplexity. 
Comparisons are made between the pre-trained, base fine-tuned (w/o \gls{dp} ), and gls{dp} fine-tuned models with $\varepsilon \in \{10,1,0.1\}$.}
  \label{fig:utility_approach}
\end{figure}
In order to investigate the effect of \gls{dp} on the performance of \gls{codellms} we take two levels of analysis: (1) Overall Code Generation Performance and (2) Performance on Fine-Tuned Dataset. To assess the overall code generation performance, we employ the HumanEval benchmark \cite{chen_evaluating_2021} benchmark. This benchmark has been widely used in the literature to evaluate the code generation capabilities of \gls{codellms}. Consequently, it functions as a standard reference point for evaluating the general impact of \gls{dp} on model utility.

Even though the HumanEval benchmark is often a good indicator of the overall code generation performance, it does not fully capture the model's ability to generate code that is specific to the fine-tuned dataset. In particular, given the potential of the noise to impact the model's ability to learn the specific patterns and structures present in the fine-tuned dataset, this approach is noteworthy. Therefore, we also evaluate the functional correctness on a HumanEval-style benchmark derived from the fine-tuning dataset itself, namely \gls{spenc}. This in-domain benchmark allows us to directly evaluate the performance of the models on tasks and structures that are represented in the fine-tuning dataset.

Finally, we compare the perplexity scores of the models on the test set of the fine-tuning dataset. This provides an additional metric to evaluate the models' performance and helps to understand the impact of \gls{dp} on the models' ability to predict and generate code based on the fine-tuned dataset. This two-level analysis enables a comprehensive evaluation of the impact of gls{dp} on the performance of \gls{codellms}, considering both their general code generation capabilities and their capacity to generate code tailored to the fine-tuned dataset.

\subsection{Energy and Efficiency Evaluation}

To cover the last research question, impact of \gls{dp} on the energy consumption and resource usage of \gls{codellms}, we use the \texttt{codecarbon} library \cite{courty_mlco2codecarbon_2024} to track the energy consumption and carbon emissions during the fine-tuning process. The library provides a simple interface to monitor the energy consumption of the hardware components, including CPU, GPU, and RAM, as well as the overall energy consumption of the system. We integrate the library into our fine-tuning pipeline to automatically log the energy consumption and carbon emissions during the training process. We then analyze the logged data to compare the energy consumption and carbon emissions of the models fine-tuned with and without \gls{dp}. We only focus on the total energy consumption (in kWh), power usage (in W), average training time per epoch (in seconds), and average throughput (in samples/second) during the fine-tuning process. With this analysis, we aim to understand the trade-offs between privacy, utility, sustainability, and resource usage in the context of \gls{dp} fine-tuning of \gls{codellms}.

\section{Results and Discussion}
Here, we present the results of our study, organized according to the \hyperref[sec:rqs]{research questions}. We first analyze the types of snippets memorized by \gls{codellms}, then evaluate the impact of \gls{dp} on memorization and code generation capabilities across different models and settings.

\subsection{RQ1: Identifying and Understanding Memorization in \gls{codellms}}
\label{rq:types}

\begin{figure}[!htbp]
  \centering
  \includegraphics[width=1\columnwidth]{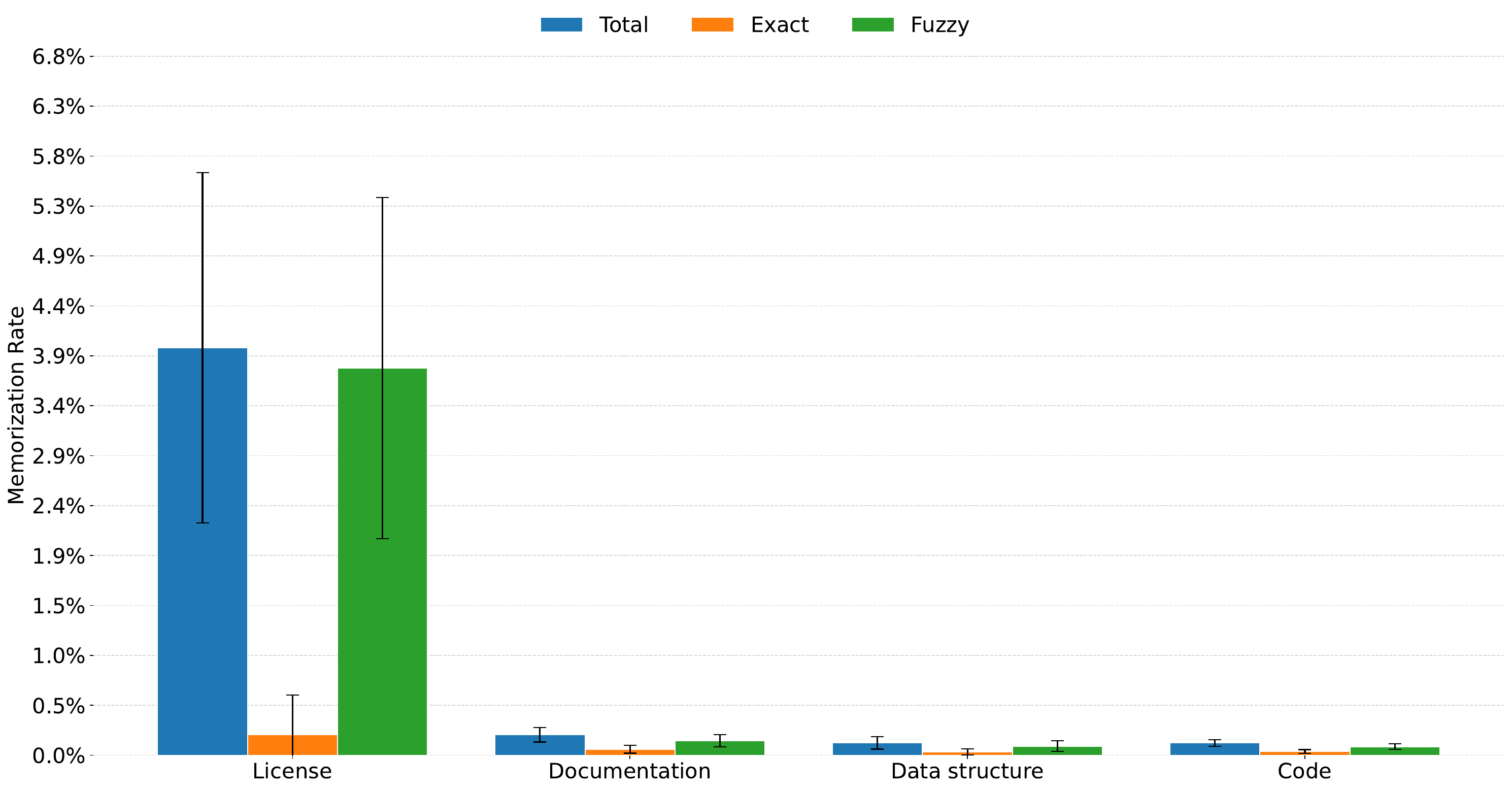}
  \caption{\textbf{Memorization rates across snippet types.}
Breakdown of memorization into \emph{exact} and \emph{fuzzy} matches, which are mutually exclusive, with their sum giving the \emph{total}.}
  \label{fig:memorization_rates_types}
\end{figure}

The first research question in this study sought to identify and understand the different types of memorization that can occur in \gls{codellms}. This involves examining the various ways in which these models can memorize training data, as well as the factors that influence this memorization behavior. To this end, we performed a data extraction attack to identify memorized snippets and analyze their characteristics.

 Regarding the categorization, our benchmark consists of four high-level categories, whose distribution is as follows: License (0.7\%), Documentation (23.6\%), Code (57.2\%), and Data Structures (18.4\%).
To understand further the specific type of code statements, 
we further refined the majority category \emph{Code} into six distinct sub-categories, whose distribution is as follows: Control Flow (25.2\%), Import Statements (17.9\%), Testing Code (23.5\%), Expressions (15.3\%), Definitions (13.4\%), and Declarations (4.5\%).

These results show that \gls{codellms} can memorize a wide range of code snippets, including those that are frequently occurring in the training data, as well as those that are more complex and less common. It is apparent from the Figure \ref{fig:memorization_rates_types} that licenses are, with a large margin, the most frequently memorized type of snippet across all models. This is likely because licenses are often included in code repositories and are therefore more likely to be present in the training data. However, this memorization behavior does not only apply to licenses, as we also observe memorization of all the other types of snippets, including documentation, data structures, and even unique code snippets. Even the memorization rates are low, the potential risks associated with this behavior are significant, particularly in terms of privacy and security. Also, it is noteworthy to remark that these memorization rates are lower bounds, as our data extraction attack does not capture all instances of memorization. And considering the large scale of the training datasets, even a small percentage of memorization can translate to a significant number of memorized snippets.

\begin{figure}[!htbp]
  \centering
  \includegraphics[width=\columnwidth]{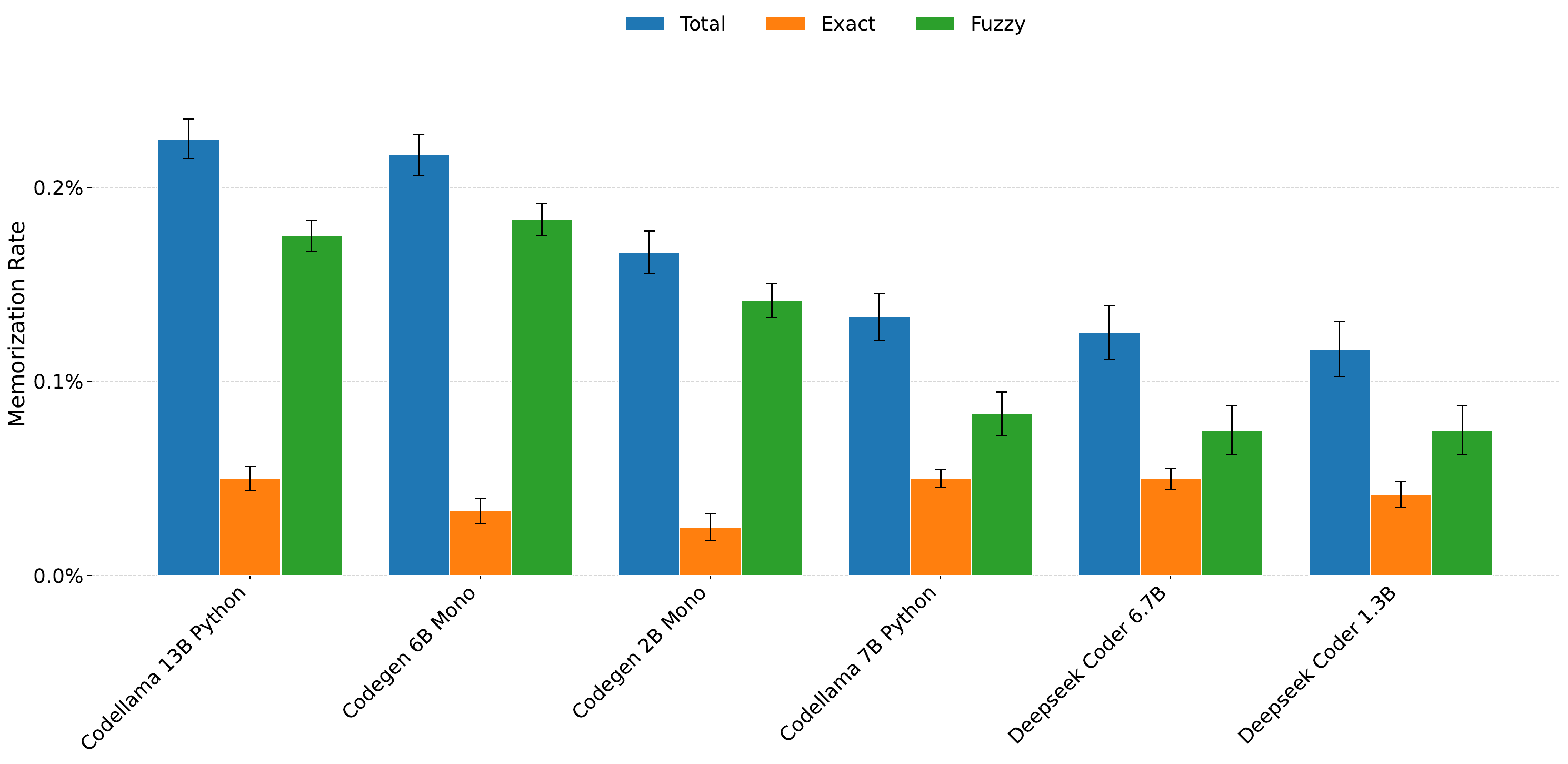}
\caption{\textbf{Model-wise memorization rates.}
Memorization is decomposed into \emph{exact} and \emph{fuzzy} matches, which are mutually exclusive, with their sum giving the \emph{total}. Memorization risk is prevalent across models, however, model choice has a significant impact on the extent of memorization.}
  \label{fig:memo_model_comp}
\end{figure}

\begin{figure}[!htbp]
  \centering
\includegraphics[width=1\columnwidth]{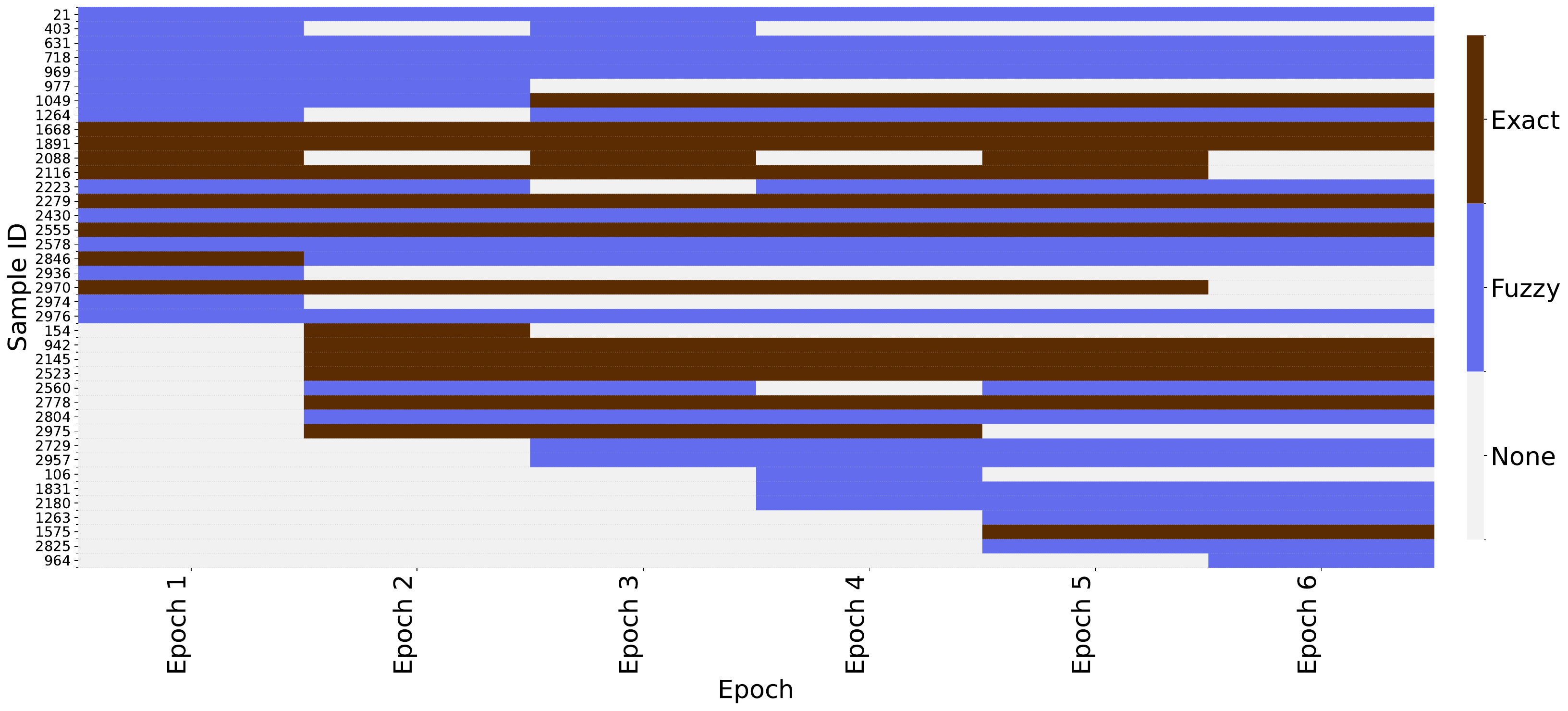}  
\caption{\textbf{Evolution of memorization during base fine-tuning (CodeLlama 7B Python).}
Trajectory of exact and fuzzy memorization across training epochs. Memorization is consolidated over time, with snippets being memorized or forgotten at different stages, but stabilizing as fine-tuning progresses. This highlights the dynamic nature of memorization in \glspl{codellms}.}
  \label{fig:memorization_order}
\end{figure}
An interesting observation from the detailed analysis of the code snippets is that control flow statements (e.g., if, for, while) are the second most commonly memorized constructs (25.2\%) across all models and fine-tuning settings. This finding suggests that business logic and decision-making processes are in danger of being memorized by \gls{codellms}, which could have serious implications for intellectual property and privacy. However, this memorization is mostly fuzzy memorization. This is also important because most of the existing works on memorization focus on exact memorization, which may underestimate the true extent of memorization in \gls{codellms}.

Following the identification of memorized snippets, we analyzed their characteristics to understand the factors that influence memorization behavior. We analyzed the frequency of the snippets in the training data, their complexity (measured in terms of gzip entropy), and their order of appearance in the fine-tuning. Our findings show that frequency has a significant impact on memorization, with more frequent snippets being more likely to be memorized (OR=3.17, p$<$0.01 for snippet types, OR=8.91, p$<$0.01 for code types). This is consistent with prior work on memorization in language models \cite{carlini_secret_2019, yang_unveiling_2024}, which has shown that models are more likely to memorize frequently occurring data. However, we also found that complexity plays a role in memorization, with more compressible, simpler snippets being more likely to be memorized (OR=0.60, p$<$0.01 for snippet types, OR=0.38, p$<$0.01 for code types). This suggests that models may be more likely to memorize snippets that are easier to learn and reproduce, which could have implications for the types of data that are included in training datasets.

Finally, as can be seen from Figure \ref{fig:memorization_order}, we observed the progressive consolidation of memorization behavior over the course of fine-tuning. In other words, a snippet can be memorized or forgotten at different stages of the fine-tuning process, but as fine-tuning progresses, the model tends to consolidate its memorization behavior, leading to a more stable set of memorized snippets. This finding highlights the dynamic nature of memorization in \gls{codellms} and suggests that the fine-tuning process itself can influence the extent and nature of memorization. 

\begin{mdframed}[style=TakeawayStyle,align=center]
  \textbf{\textcolor{accentDark}{Takeaway:}}  Memorization in \gls{codellms} is most pronounced for frequent and simple snippets, increases with model size, and evolves through a progressive consolidation process during fine-tuning.
\end{mdframed}

\subsection{RQ2: Effect of \gls{dp} on Mitigating Memorization in \gls{codellms}}
\begin{figure}[!htbp]
  \centering
\includegraphics[width=0.6\columnwidth]{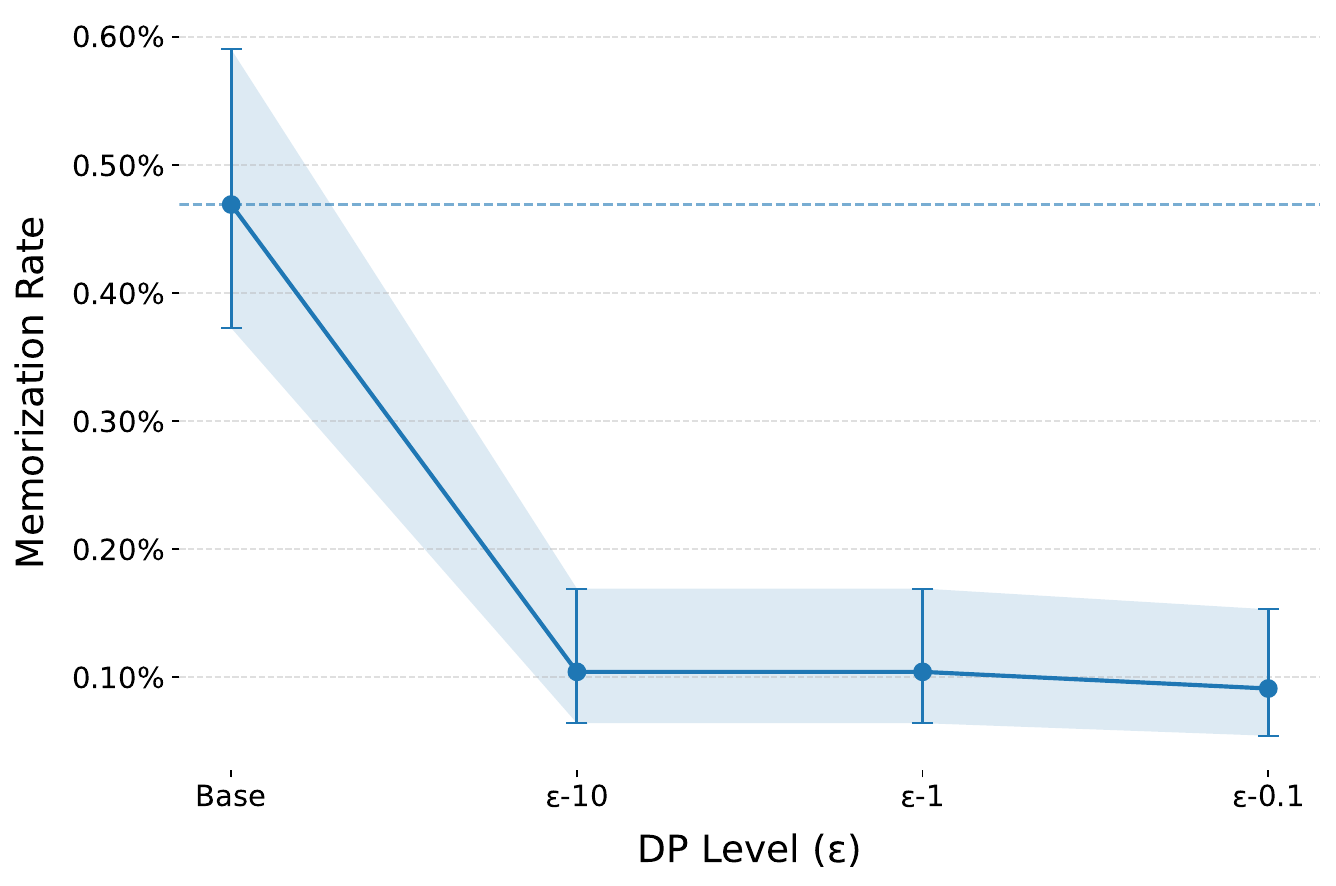}  
\caption{\textbf{Average effect of \gls{dp} on memorization across models.}
Applying \gls{dp} significantly decreases the need for memorization compared to the non-DP baseline. Even mild privacy settings ($\varepsilon = 10$) sharply reduce memorization rates, and stronger settings ($\varepsilon = 1, 0.1$) provide additional reductions, though the differences are minimal.}
  \label{fig:avg_memo_dp_change}
\end{figure}

The second research question was designed to evaluate the effectiveness of \gls{dp} in mitigating memorization risks in \gls{codellms}, as well as to investigate whether the effectiveness of \gls{dp} varies across different types of memorized snippets. To this end, we fine-tuned multiple \gls{codellms} with and without \gls{dp} and performed a data extraction attack to identify memorized snippets in each model. We then compared the number and types of memorized snippets across models to assess the impact of \gls{dp} on memorization.

\begin{figure}[!htbp]
  \centering
  \begin{subfigure}{0.48\columnwidth}
    \centering
    \includegraphics[width=\linewidth]{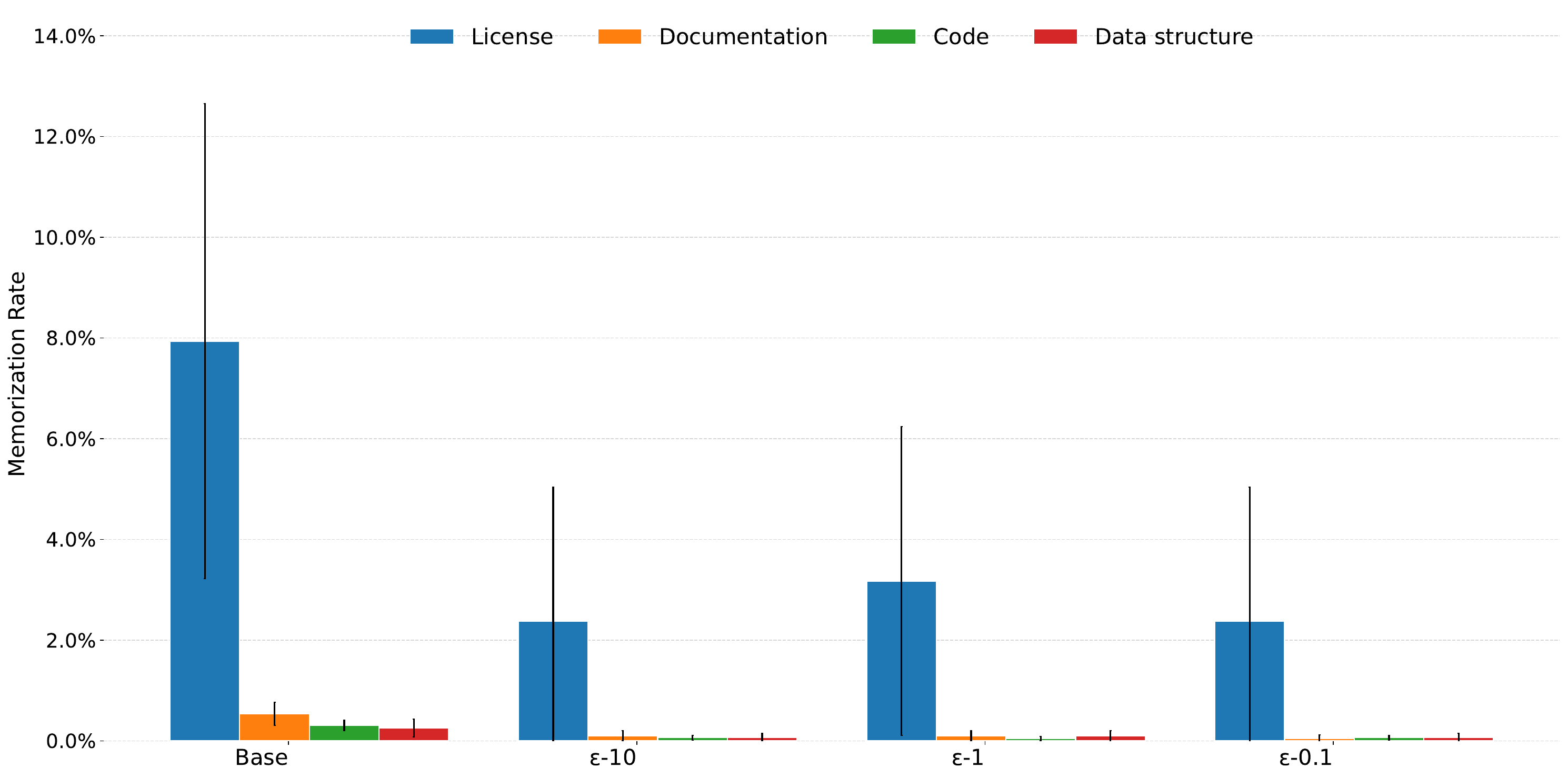}
    \caption{High-level snippet types, and how their memorization rate changes with \gls{dp}.}
    \label{subfig:high_level}
  \end{subfigure}\hfill
  \begin{subfigure}{0.48\columnwidth}
    \centering
    \includegraphics[width=\linewidth]{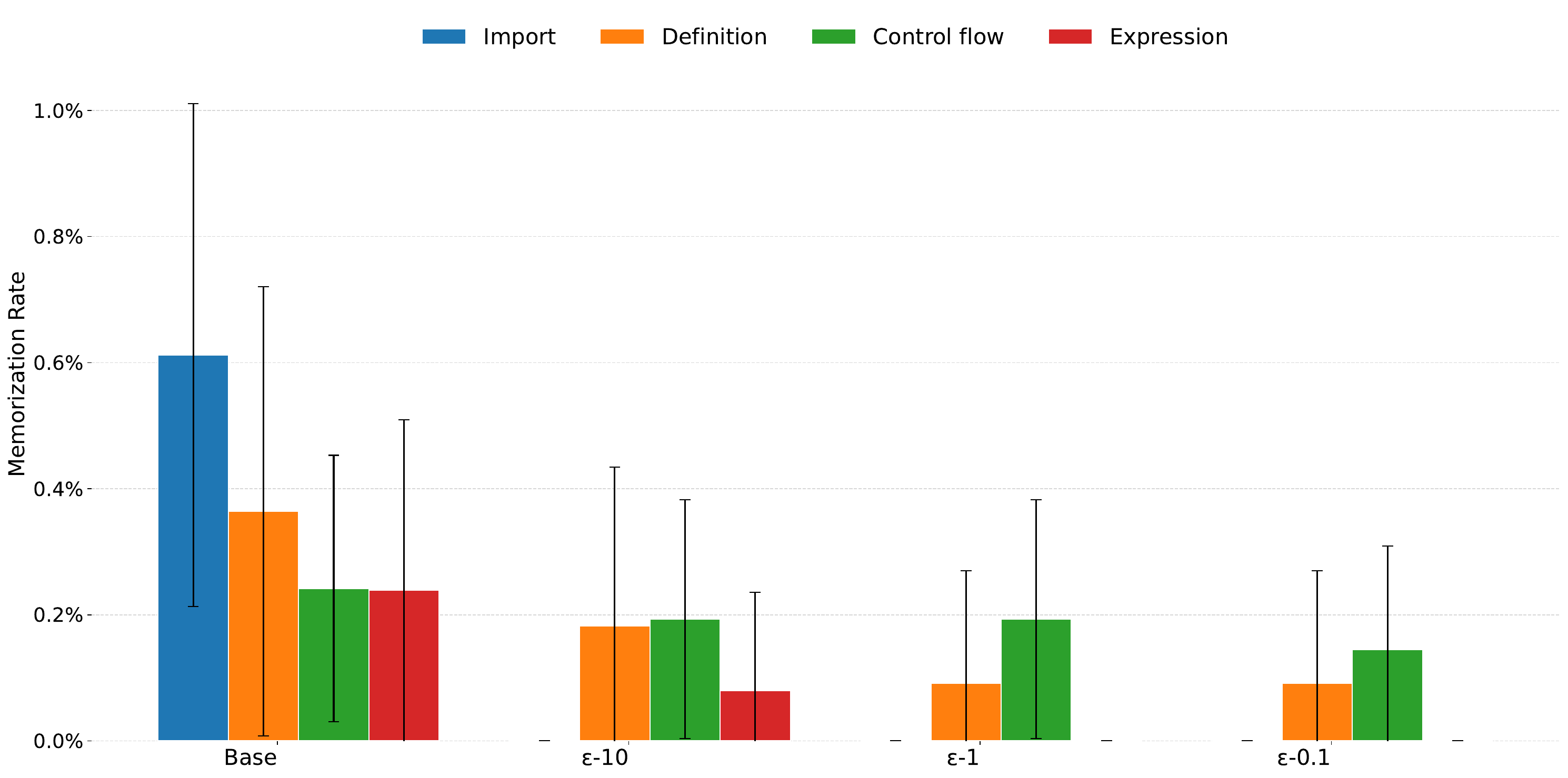}
    \caption{Lower-level code categories, and how their memorization rate changes with \gls{dp}.}
    \label{subfig:low_level}
  \end{subfigure}
  \caption{\gls{dp} impact on memorization by snippet category: (a) high-level snippet types; (b) lower-level code categories.}
  \label{fig:high_level_ol}
\end{figure}

The results highlight the effectiveness of \gls{dp} in reducing memorization risks in \gls{codellms}. As shown in Figure \ref{fig:avg_memo_dp_change}, models fine-tuned with \gls{dp} exhibit a significant reduction in the number of memorized snippets compared to their non-\gls{dp} counterparts. A repeated-measures ANOVA confirmed a strong overall effect of \textit{epsilon} on memorization \(F(3, 15) = 36.09,\; p < .001\). Pairwise Wilcoxon tests further showed that models trained with \gls{dp} at lower \textit{epsilon} values (0.1, 1.0) memorized significantly less than their non-\gls{dp} baseline counterparts (p = 0.031), whereas differences among \gls{dp} settings were not significant. Another key finding is that the effectiveness of \gls{dp} in mitigating memorization risks does not vary significantly across different types of memorized snippets. As illustrated in Figure \ref{fig:memorization_rates_types}, the reduction in memorization is consistent across all snippet types, including licenses, documentation, data structures, and unique code snippets. This suggests that \gls{dp} provides a robust defense against memorization risks, regardless of the specific characteristics of the memorized data. For instance, licenses, which are the most frequently memorized type of snippet, still remain the most memorized overall, but their memorization rate drops by about 70\% when even a slight amount of noise is injected ($\epsilon = 10$). This mitigation is even more pronounced for other types of snippets, such as import statements. As shown in Figure \ref{subfig:low_level}, import statements are by far the most memorized type of snippet in the non-DP setting, but their memorization rate drops by about 100\% when even a slight amount of noise is injected ($\epsilon = 10$). On the hand, more complex structures, such as control flow statements, see a more modest reduction in memorization, with a drop of about 20\% at the same privacy level. This indicates that while \gls{dp} is effective across the board, its impact may be more pronounced for certain types of snippets, particularly those that are simpler and more repetitive in nature.

\begin{mdframed}[style=TakeawayStyle,align=center]
  \textbf{\textcolor{accentDark}{Takeaway:}}  \gls{dp} substantially reduces memorization in \gls{codellms} across all the tested snippet types. The snippet types most prone to memorization are also the most effectively mitigated by \gls{dp}. 
\end{mdframed}

\subsection{RQ3: Effect of \gls{dp} on Code Generation Capabilities}

\begin{figure}[!htbp]
  \centering
\includegraphics[width=1\columnwidth]{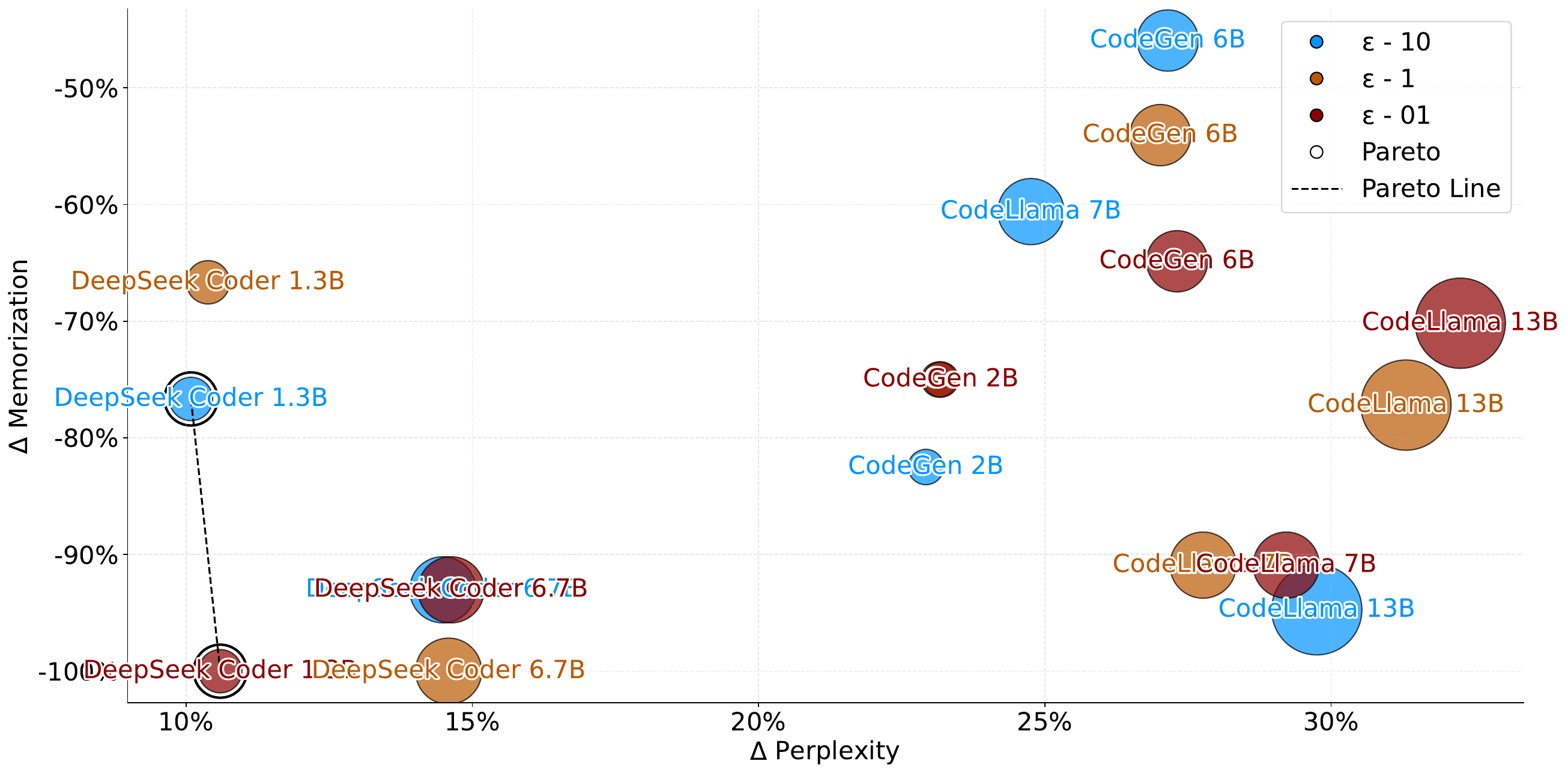}  
  \caption{\textbf{Privacy–utility trade-off in memorization and perplexity.}
  $\Delta$ Perplexity (lower is better) vs.\ $\Delta$ Memorization (higher is better), where each point represents the change relative to the non-DP baseline for the corresponding model. Points are colored by $\varepsilon\!\in\!\{10,1,0.1\}$; the dashed line marks Pareto-efficient models.}
  \label{fig:perplexity_memo_change_model_wise}
\end{figure}

\begin{figure}[!htbp]
  \centering
\includegraphics[width=1\columnwidth]{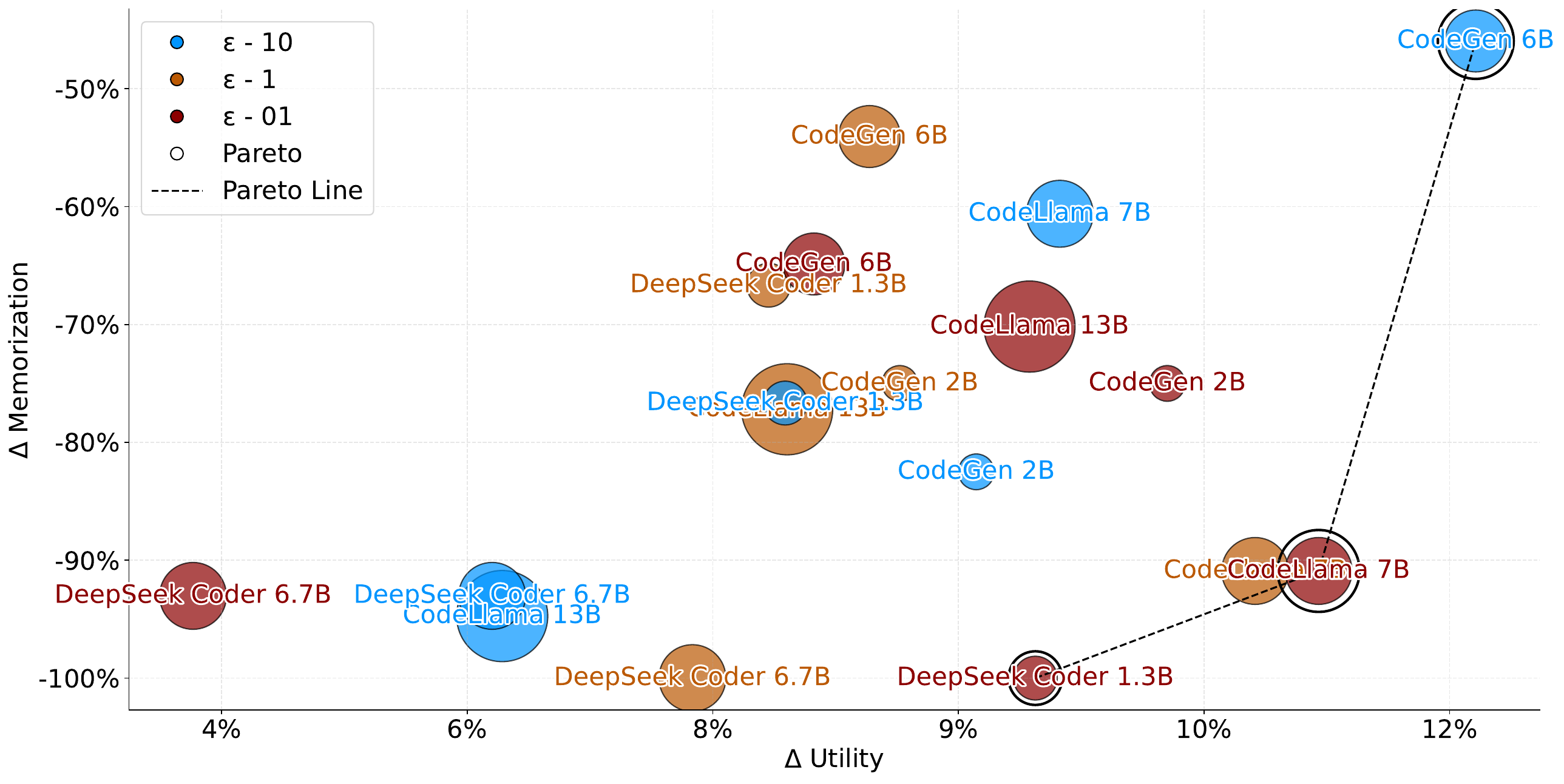}  
  \caption{\textbf{Privacy–utility trade-off in functional correctness.}
  $\Delta$ Utility (average change in HumanEval and SPE-NC functional correctness; higher is better) vs.\ $\Delta$ Memorization (higher is better), where each point represents the change relative to the non-DP baseline for the corresponding model. Points are colored by $\varepsilon\!\in\!\{10,1,0.1\}$; the dashed line marks Pareto-efficient models.}
  \label{fig:utility_memo_change_model_wise}
\end{figure}

The third research question focuses on evaluating the impact of \gls{dp} on the code generation capabilities of \gls{codellms}. To this end, we fine-tuned multiple \gls{codellms} with and without \gls{dp} and evaluated their performance using two benchmarks: HumanEval and \gls{spenc}. The HumanEval benchmark assesses the overall code generation performance, while the \gls{spenc} benchmark evaluates the in-domain performance on tasks and structures represented in the fine-tuning dataset. Additionally, we compared the perplexity scores of the models on the test set of the fine-tuning dataset to provide an additional metric for evaluating model performance.

\begin{figure}[!htbp]
  \centering
\includegraphics[width=0.8\columnwidth]{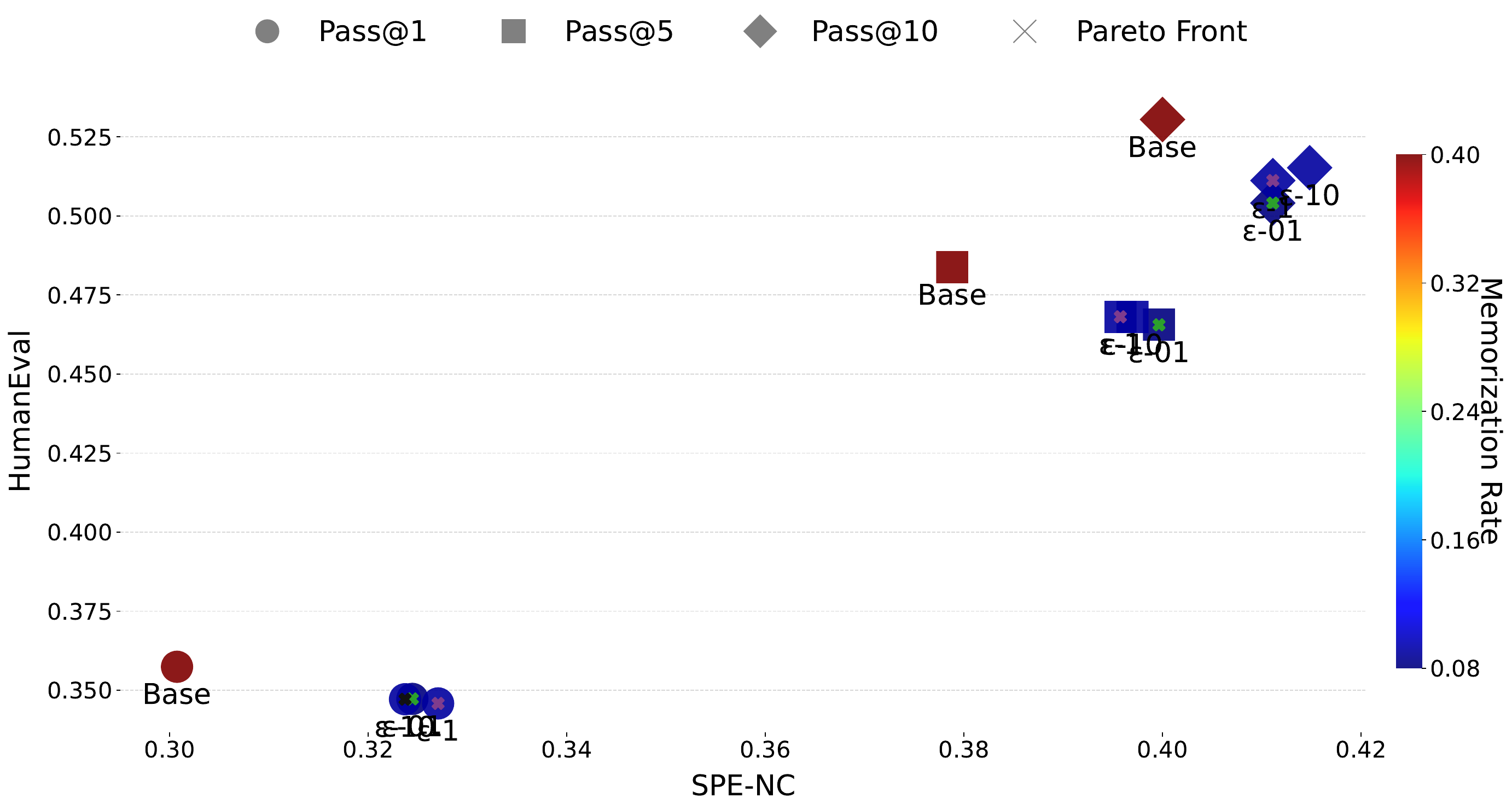}  
\caption{\textbf{Comparison of general and fine-tuning specific functional correctness with memorization.}
Performance on HumanEval (general benchmark) and SPE-NC (fine-tuning dataset benchmark) is shown under different fine-tuning settings (base, non-DP, and DP with $\varepsilon \in \{10,1,0.1\}$). 
Marker shapes represent pass@1, pass@5, and pass@10, while color encodes memorization rate. 
Results highlight the trade-off between functional correctness and memorization, with DP fine-tuning reducing memorization while maintaining competitive utility.}
  \label{fig:memo_humaneval_spe}
\end{figure}

Starting with the HumanEval benchmark, our results show that the application of \gls{dp} does not significantly degrade the overall code generation performance of \gls{codellms}. As shown in Figure \ref{fig:memo_humaneval_spe}, models fine-tuned with \gls{dp} achieve comparable pass@k scores to their non-\gls{dp} counterparts across all three epsilon values and k values (1, 5, and 10). Pairwise Wilcoxon tests further showed no significant differences between any of the \gls{dp} settings and the non-\gls{dp} baseline (all p > 0.05). This suggests that \gls{dp} does not necessarily compromise the model's overall code generation capabilities, even at lower privacy levels.

Moving to the in-domain performance on the \gls{spenc} benchmark, we observe a similar trend. As illustrated in Figure \ref{fig:utility_memo_change_model_wise}, models fine-tuned with \gls{dp} achieve pass@k scores that are on par with, or even slightly better than, their non-\gls{dp} counterparts across all three epsilon values and k values (1, 5, and 10). Once again, pairwise Wilcoxon tests showed no significant differences between any of the \gls{dp} settings and the non-\gls{dp} baseline (all p > 0.05). This indicates that \gls{dp} does not hinder the model's ability to learn and generate code that is specific to the fine-tuning dataset.

However, when examining the perplexity scores on the test set of the fine-tuning dataset, we find that models fine-tuned with \gls{dp} exhibit slightly higher perplexity scores compared to their non-\gls{dp} counterparts, and increasing privacy levels (lower epsilon values) lead to higher perplexity scores. Pairwise Wilcoxon tests confirmed that models trained with \gls{dp} at lower \textit{epsilon} values (0.1, 1.0) had significantly higher perplexity than their non-\gls{dp} baseline counterparts (p = 0.031). This worsening perplexity is further consistent within \gls{dp} settings, where lower epsilon values correspond to higher perplexity scores. The only exception is the $\epsilon = 10$ setting, which shows no significant difference from the $\epsilon = 1$ setting (p = 0.093). This suggests that while \gls{dp} may introduce some noise that affects the model's ability to predict and generate code based on the fine-tuned dataset, this does not necessarily translate to a degradation in code generation performance as measured by pass@k scores on the HumanEval and \gls{spenc} benchmarks. Thus, it is important to consider multiple metrics when evaluating the impact of \gls{dp} on \gls{codellms}, as different metrics may capture different aspects of model performance.

\begin{mdframed}[style=TakeawayStyle,align=center]
  \textbf{\textcolor{accentDark}{Takeaway:}}  \gls{dp} slightly increases perplexity but preserves, and can even enhance, the code generation capabilities of \gls{codellms}, thus making it feasible to apply \gls{dp} in practice without significantly compromising model utility.
\end{mdframed}

\subsection{RQ4: Effect of \gls{dp} on Efficiency and Energy Consumption}

\begin{figure}[!htbp]
  \centering
\includegraphics[width=1\columnwidth]{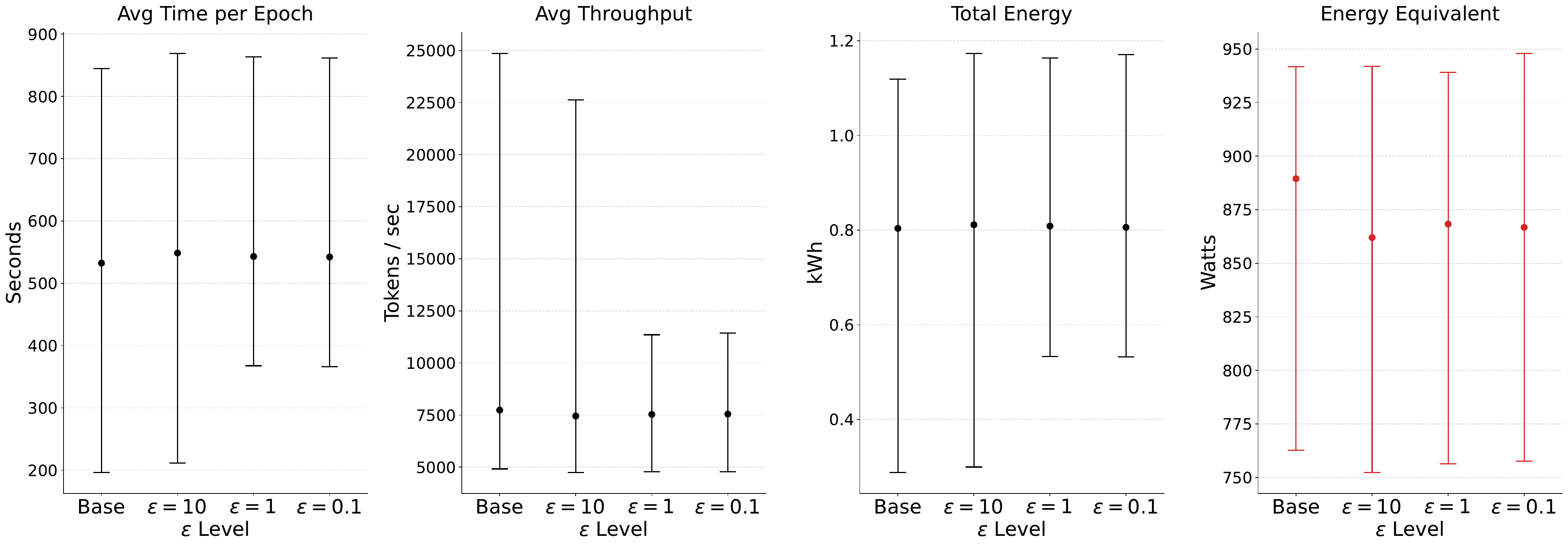}  
\caption{\textbf{Impact of differential privacy on energy and efficiency.}
Average time per epoch, throughput (tokens/sec), total energy consumption (kWh), and energy equivalent (Watts) for the baseline (non-DP) and \gls{dp} levels ($\varepsilon \in \{10,1,0.1\}$).}
  \label{fig:energy_efficiency_results}
\end{figure}

The purpose of the fourth research question was to investigate the impact of \gls{dp} on the training efficiency and energy consumption of \gls{codellms}, as well as to explore whether this impact varies with the privacy level. To achieve this, we tracked the energy consumption and training time during the fine-tuning process of multiple \gls{codellms} with and without \gls{dp}. We then compared the energy consumption, training time, and throughput across models to assess the impact of \gls{dp} on efficiency and resource usage.

The Figure \ref{fig:energy_efficiency_results} shows the results of our analysis. We observe that the application of \gls{dp} leads to an increase in energy consumption and training time compared to the non-\gls{dp} baseline. A repeated-measures ANOVA confirmed that the differences in energy consumption ($F(3,15)=1.35, p=0.30$), power usage ($F(3,15)=1.45, p=0.27$), average training time per epoch ($F(3,15)=1.41, p=0.28$), and average throughput ($F(3,15)=1.07, p=0.39$) are statistically insignificant, indicating that \gls{dp} can be implemented in \gls{codellms} without incurring meaningful overhead. This finding underscores the practicality of adopting \gls{dp} in real-world training pipelines, as privacy can be enhanced without compromising efficiency or sustainability.

\begin{mdframed}[style=TakeawayStyle,align=center]
  \textbf{\textcolor{accentDark}{Takeaway:}}  \gls{dp} can be a practical and sustainable approach for enhancing privacy in \gls{codellms}, as it does not introduce significant overhead in terms of energy consumption and efficiency, even at varying privacy levels.
\end{mdframed}

\section{Threats to Validity}\label{sec:threats-to-validity}

This study has several limitations that may affect the interpretation and generalizability of the results. First, our analysis is based on a specific set of \gls{codellms} and fine-tuning datasets, which may not be representative of all models and datasets used in practice. Future work could explore a broader range of models and datasets to validate the findings. Second, the data extraction attack employed in this study may not capture all instances of memorization, particularly fuzzy memorization. More sophisticated attack methods could be developed to better identify and quantify memorization in \gls{codellms}. Third, the evaluation metrics used in this study, such as functional correctness and perplexity, may not fully capture the utility of \gls{codellms} in real-world scenarios. Future research could consider additional metrics that better reflect the practical utility of these models. Finally, the implementation of \gls{dp} in this study is based on specific algorithms and parameters, which may not be optimal for all scenarios. Further research could investigate alternative \gls{dp} techniques and configurations to better understand their impact on memorization and utility in \gls{codellms}.

\section{Conclusion and Future Work}\label{sec:conclusion-and-future-work}
The main purpose of this study was to investigate the effectiveness of \gls{dp} in mitigating memorization risks in \gls{codellms}, while also evaluating its impact on code generation capabilities and training efficiency. To systematically address these objectives, we formulated four research questions (RQs) that guided our experimental design and analysis. We first sought to identify and understand the categories of code snippets that are most prone to memorization by \gls{codellms} and the factors driving that memorization (RQ1). Building on this, we then evaluated the effectiveness of \gls{dp} in mitigating these memorization risks and whether its effectiveness varied across different types of memorized snippets (RQ2). Next, we assessed how \gls{dp} impacts the code generation capabilities of \gls{codellms}, focusing on both syntactic and semantic correctness (RQ3). Finally, we examined the impact of \gls{dp} on training efficiency and energy consumption, considering how these factors vary with different privacy levels (RQ4). By addressing these research questions, we aimed to provide a comprehensive understanding of the trade-offs involved in applying \gls{dp} to \gls{codellms}, balancing privacy, utility, and resource usage to have a privacy-preserving yet high-performing \gls{codellms}. 

Our findings show that frequently occurring and simpler code snippets are more prone to memorization by \gls{codellms}, with licenses being the most frequently memorized type of snippet. We also found that \gls{dp} is effective in mitigating memorization risks across all types of snippets, without a degradation in code generation capabilities and efficiency and energy consumption. These results highlight the potential of \gls{dp} as a robust defense mechanism against memorization risks in \gls{codellms}, while also maintaining their utility and efficiency.

To the best of our knowledge, this is the first comprehensive study that systematically evaluates the effectiveness of \gls{dp} in \gls{codellms}. We believe that the increasing demand for privacy-preserving yet high-performing \gls{codellms} will further motivate research on balancing memorization risks with utility, opening new directions for model training, evaluation, and deployment in sensitive domains. Our findings suggest that \gls{dp} can be a valuable tool in this regard, providing strong privacy guarantees without compromising the performance of \gls{codellms}. Future work could explore the integration of \gls{dp} with other privacy-preserving techniques, such as federated learning or secure multi-party computation, to further enhance the privacy and security of \gls{codellms}. Additionally, investigating the impact of \gls{dp} on other aspects of \gls{codellms}, such as code comprehension and debugging capabilities, could provide a more holistic understanding of its effects on model performance. And finally, category-aware \gls{dp} mechanisms that tailor the privacy guarantees based on the characteristics of the code snippets could be developed to optimize the trade-off between privacy and utility in \gls{codellms}. 

\section{Data Availability.}
All data, and the scripts used for analysis are available in our replication package~\cite{reppackage}.

\bibliographystyle{ACM-Reference-Format}

\bibliography{references,references-2}

\newpage
\onecolumn

\end{document}